\documentclass[]{spie}  
\usepackage[]{graphicx}

\title{The Robo-AO software: Fully autonomous operation of a laser guide star adaptive optics and science system} 

\author{Reed L. Riddle\supit{a}, Mahesh P. Burse\supit{b}, Nicholas M. Law\supit{c}, Shriharsh P. Tendulkar\supit{a}, Christoph Baranec\supit{a}, Alexander R. Rudy\supit{d}, Marland Sitt\supit{e}, Ankit Arya\supit{f}, Athanasios Papadopoulos\supit{g}, A. N. Ramaprakash\supit{b}, and Richard G. Dekany\supit{a}
\skiplinehalf
\supit{a}Caltech Optical Observatories, California Institute of Technology, 1200 E. California blvd., MC 11-17, Pasadena, CA, 91125, USA;  
\supit{b}Inter-University Centre for Astronomy and Astrophysics, Post Bag 4, Ganeshkhind, Pune, India 411 007;  
\supit{c}Dunlap Institute for Astronomy and Astrophysics, University of Toronto, 50 St. George Street, Toronto M5S 3H4, Ontario, Canada;  
\supit{d}National Central University, No.300, Jhongda Rd., Jhongli City, Taoyuan County 32001, Taiwan (R.O.C.);  
\supit{e}Stanford University, 450 Serra Mall, Stanford, CA 94305, USA;  
\supit{f}Mississippi State University, 103 Rl Jones Cir  Starkville, MS 39759, USA;  
\supit{g}Aristotle University of Thessaloniki, 541 24, Greece
}

\authorinfo{Send correspondence to Reed L. Riddle. E-mail: riddle@caltech.edu, Telephone: 1 626 429 8420\\  
}

\begin{document} 
  \maketitle 

\begin{abstract}
Robo-AO is the first astronomical laser guide star adaptive optics (AO) system designed to operate completely independent of human supervision. A single computer commands the AO system, the laser guide star, visible and near-infrared science cameras (which double as tip-tip sensors), the telescope, and other instrument functions. Autonomous startup and shutdown sequences as well as concatenated visible observations were demonstrated in late 2011. The fully robotic software is currently operating during a month long demonstration of Robo-AO at the Palomar Observatory 60-inch telescope.\end{abstract}



\section{INTRODUCTION}
\label{sec:intro}  

Robo-AO is an autonomous adaptive optics instrument that robotically operates a telescope and laser guide star and science system to observe several different classes of astronomical objects\cite{Baranec}.  It is the first system that operates a laser guide star without human oversight, and can operate continuously and robotically.  The software architecture for the Robo-AO system has been designed to be as robust as possible, but also as a system that is simple and flexible to manage and operate.  Since Robo-AO is a robotic instrument, it  functions independent of human interaction.  Therefore, the computer code has been created in such a way that it is tolerant to failure of a function, and is fail safe (i.e. if something untoward happens the system will shut down into a state where the telescope and instrument are in no danger).  This required a level of software development that is meticulous, that takes into account many possible failures and mitigation procedures in the software development phase, and that is as immune as possible to random errors, bugs and crashes.  The Robo-AO software must also be flexible in design, so that separate parts can be easily combined to create the entire operating system.   

\section{SOFTWARE ARCHITECTURE}

\subsection{System Basics}

The Robo-AO computer uses Fedora 13 as the base operating system. The base Fedora installation is expanded to include the development environment and several packages to support basic functions such as FITS file management and the hardware drivers.  The system does not use a real time kernel; this choice was made to save on complication and increase portability of the software.  In practice, the operation software does not require better than microsecond timing, and the Linux operating system can handle that with ease.

All source code for the Robo-AO project is written in C++.  At the time of this writing, the software consists of almost 100,000 lines of documented source code.  Scripting for some minor housekeeping functions is done using the Bash shell scripting language; all configuration files that are read by the Robo-AO system are in Bash shell formatted text files.  

A threaded logging system was created for Robo-AO when other solutions were deemed not quite flexible enough for our purposes.  Telemetry is handled via text file output; both low (i.e. ~1Hz) and high (i.e. at the rate of the fastest hardware systems) speed telemetry systems are available.  A future upgrade path for the system is a telemetry database.  A web monitoring interface is available, using PHP for the basic structure, with other software (for making plots or parsing data) using Javascript.  There is a simple text menu based interface for manual control of the system; since Robo-AO is a robotic system a GUI is not required.

A multithreaded FITS handling system, based on the C++ version of CFITSIO, CCFITS, was built to handle the Robo-AO data output.  The FITS system can create single images and data cubes, and includes an extensive collection of header parameters that detail the information required to analyze the data as well as the performance of the instrument system.

Communications between the subsystems of the Robo-AO control software use a custom TCP/IP protocol created for the instrument.  This protocol is used to pass commands and exchange telemetry between each of the subsystems, and automatically detects when communication is dropped and restarts the subsystem that is unresponsive.  

\subsection{Hardware Control Software}

The Robo-AO hardware interface software has been developed as a modular system.  The software to control each hardware subsystem was developed as a set of individual modules; small standalone test programs have been created to test each of the hardware interfaces.  This modular design allows the individual subsystems to be stacked together into larger modules, which can then be managed by other facets of the robotic control system. 

The Robo-AO software controls the following hardware:

\begin{itemize}
\item Wavefront Sensor (WFS) camera
\item Deformable Mirror (DM)
\item Tip-Tilt Mirror
\item Tip-Tilt camera
\item Laser system (laser, laser chiller, safety shutter, beam steering mirror, range gate) 
\item Atmospheric dispersion correction system
\item Visible Camera
\item Infrared detector
\item Filter wheels  
\item Network Power Switch (NPS)
\item Telescope operations and monitoring
\end{itemize}

Each of the hardware interface modules handles configuration file interactions, error control for the hardware interaction, initialization and configuration, and any other interactions necessary for the specific piece of hardware.

\subsection{Subsystem Control Daemons}

The Robo-AO hardware systems are controlled by several software subsystems.  These subsystems are run as daemons in the operating system; each separately manages the hardware under its control and runs a status monitor to sample subsystem performance and register errors that occur.  

The Robo-AO subsystem daemons are:

   \begin{figure}
   \begin{center}
   \begin{tabular}{c}
   \includegraphics[height=10.7cm]{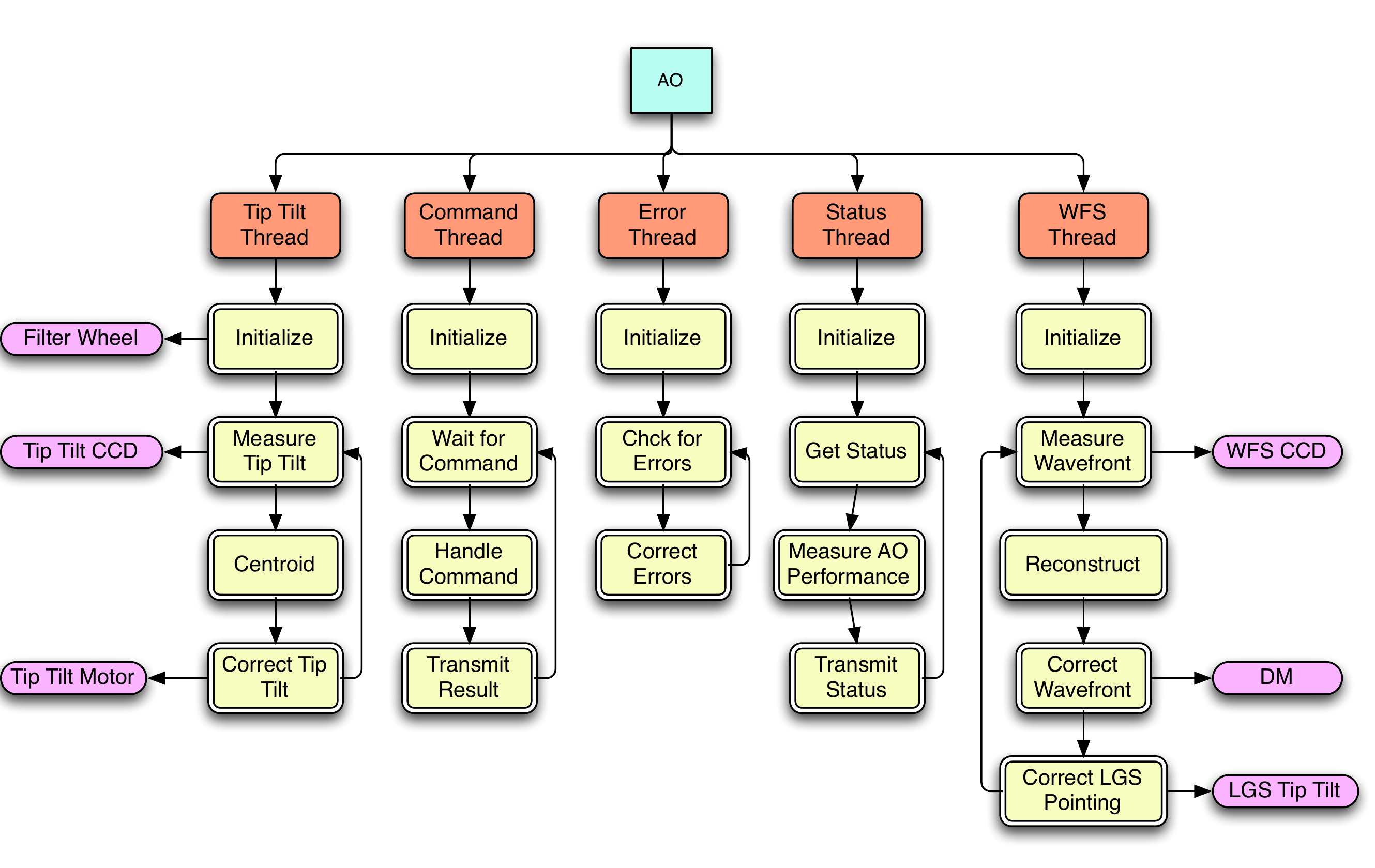}
   \end{tabular}
   \end{center}
   \caption[aosys] 
   { \label{fig:aosys} The multithreaded structure of the AO subsystem daemon.  Yellow boxes are representations of operations (which may include several subroutines), pink boxes are hardware interfaces.  }
   \end{figure} 

\begin{description}
\item [Adaptive Optics (AO)]:  Controls all of the adaptive optics functions, including the WFS CCD, DM, tip tilt system, and the laser beam steering mirror.
\item [Laser Guide Star (LGS)]:  Controls the laser system hardware operation.  Also manages the laser safety system, including automated predictive avoidance of satellites as coordinated with USSTRATCOM.
\item [Atmospheric Dispersion Corrector (ADC)]:  Operates the subsystem that corrects for atmospheric dispersion in Robo-AO imagery.
\item [Visible Instrument Camera (VIC)]:  Operates the visible science system, including the detector and filter wheels.
\item [InfraRed Camera (IRC)]:  Operates the infrared science system, including the detector and filter wheels.
\item [Telescope Control (TCS)]:  Interfaces with the telescope system to command telescope operations.
\item [Telescope Status]:  Monitors telescope status.
\item [Weather] Gathers weather condition information.
\end{description}

Each of the subsystems is composed of many separate functions that initialize the hardware, monitor its function and manage the operation of the hardware to achieve successful scientific output.  As an example, Figure~\ref{fig:aosys} shows a structural layout of the AO subsystem daemon.  Two threads independently control hardware operations as well as measurements based on the output of the two CCD cameras.  A status thread monitors the variables and hardware output to ensure that the system is functioning correctly.  A command thread accepts commands over the TCP/IP interface, executes the commands, and returns the output of the commands to the calling subsystem.  The error thread captures and corrects error states, up to and including restarting the entire subsystem hardware if a serious enough error is detected.

In essence, each of the subsystem daemons are individual robotic programs that manage their hardware and operate according to external commands.

\section{THE ROBO-AO AUTOMATION SYSTEM}

   \begin{figure}
   \begin{center}
   \begin{tabular}{c}
   \includegraphics[height=12cm]{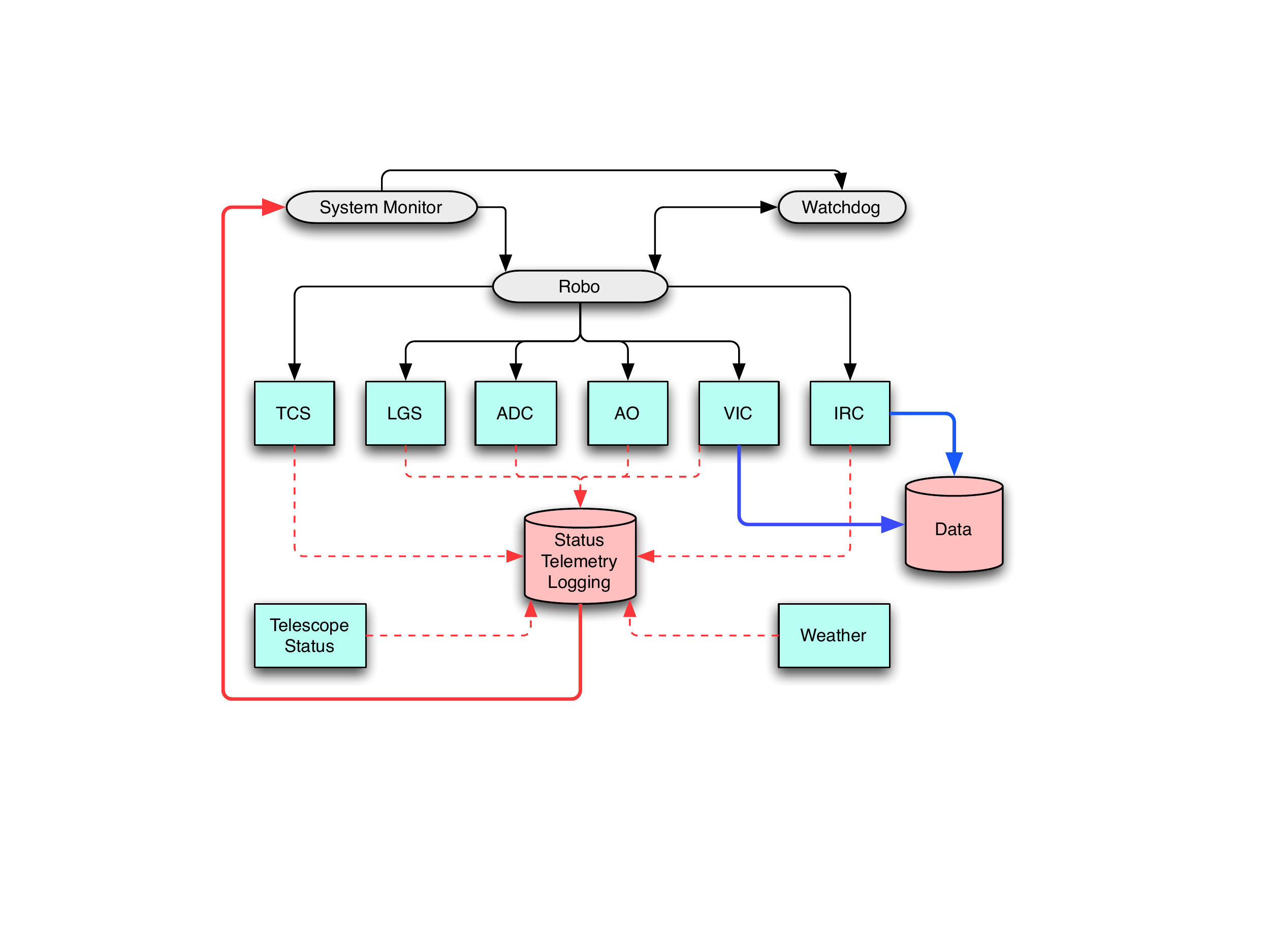}
   \end{tabular}
   \end{center}
   \caption[functions] 
   { \label{fig:functions} The automation software architecture.  Blue boxes are the hardware control subsystem daemons, gray boxes are control or oversight daemons, and red boxes are data file storage.  Red lines with arrows show the paths for telemetry through the operating system, black the command paths, and blue the data paths.}
   \end{figure} 

Figure~\ref{fig:functions} shows the overall architecture of the entire Robo-AO automated control system.  The subsystem daemons communicate their state through the TCP/IP protocol to a system monitoring service, which is used by the robotic system to control the subsystems and correct for errors.  The robotic system schedules observations and operates the instrumentation to gather the data, and a watchdog process monitors the system status and robotic system in case of errors that the robotic system misses or cannot handle.  

\subsection{System Monitor}

The System Monitor manages the information flow of the status of the subsystems to the entire robotic system.  This part of the software regularly examines the status of each of the other software elements for their state of operation.  This detects when one of the software subsystems has an error, crashes, or other issues that might hinder the proper operation of the system.  Issues are flagged and stop the operation of the automated system observations until the subsystem daemon can clear the issue.  If the subsystem cannot correct the error, the automation system can take steps, up to and including restarting subsystems, in an attempt to continue operations.  If it is unable to restart the system, it shuts everything down, leaving the system in a safe state, and sends a message for human assistance.  

\subsection{Observing Sequence}

The central engine in the automation system is the Observing Sequencing System (OSS).  This software is the conductor, controlling the operations of all of the Robo-AO subsystems to gather scientific data.  It is a complicated piece of software that has to schedule operations of several subsystems, monitor their status for errors, correctly take data that is scientifically valuable, and do it completely automatically throughout a night.  

The automation system contains parallel threads that control communications with each of the subsystems, as well as other threads that monitor dome position, subsystem errors, system health and other functions.  Communication between the threads has to be finely balanced to allow the system to function properly; it is quite easy to put multithreaded software into states where the threads become unresponsive.  Each step in the observing process has been created in a way that allows for maximum efficiency; the system can move the telescope, set up cameras, move filters, move the dome, align optics, and do all of these things at the same time to increase the rapidity of observations, all the while monitoring the health and operation of the subsystems.  This threaded architecture is key to the ability of Robo-AO to operate efficient and safely.

\subsection{System Watchdog}

The sole function of this system will be to make sure the observing system remains in operation.  If any system crashes or stops working properly, it will attempt to restart it, as well as stop telescope and laser operations immediately.  If the process cannot be restarted, then the entire telescope and instrument system will be shut down, leaving the system in a safe state, and send a message for human assistance.  This part of the system is still under development.

\subsection{Operation of the Software}

At boot, the computer starts up the supervisor tasks, which each assess the state of the hardware and software of the Robo-AO instrument.  Once it is determined that all components are in a safe state, the system begins gathering environmental data, synchronizing clocks and other functions required to start observational operations.  The robotic system remains in this state until the ephemeris determines that it has reached the time that the system starts the observation process.

\begin{figure}
   \begin{center}
   \begin{tabular}{c}
   \includegraphics[height=20.5cm]{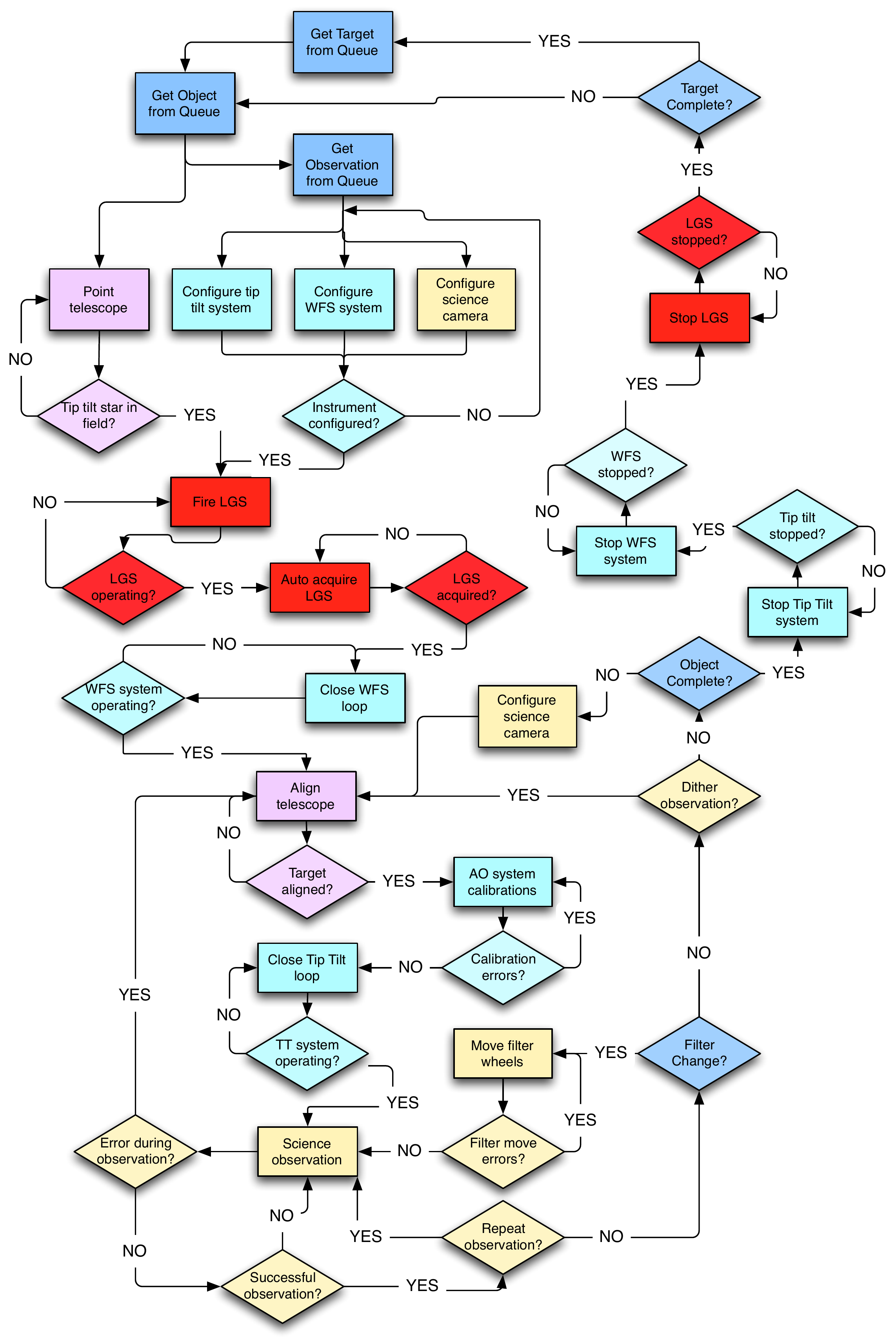}
   \end{tabular}
   \end{center}
   \caption[flowchart] 
   { \label{fig:flowchart} A flow chart demonstrating the operation of the Robo-AO robotic sequencing system.}
   \end{figure}

The initial stages of the observations are simple functions such as opening the dome to equalize the temperature and taking reference frames (i.e. dark, bias flat, and other required calibration frames) with the detectors.  Once the sky is dark enough, observations of astronomical targets commences.  The robotic observation system selects a target (initially using a simple system with the goal of developing a complete queue system), determines what types of observations are required for that particular target, slews the telescope into position, control the WFS and LGS systems to start their operation, and completes the science observations.  Figure~\ref{fig:flowchart} shows a flow chart for the operation of the Robo-AO observing system; there are many steps required to propagate the laser, operate the AO system, take science data, and each requires checks to ensure that the system is operating properly.

During an observation, the status of the subsystems is continually be monitored for errors; any errors are flagged and handled appropriately.  The environment in and around the telescope is also monitored.  Weather limits are set for the observatory (based on the safety of the equipment); if the weather should exceed these limits, the dome closes and the system goes into a safe state.  The weather monitoring continues, and when it is again safe the observation sequence begins again.  

The system continues to observe targets throughout the night, moving from one target to another on the observation list.  At the end of the night, predetermined by an ephemeris, the Robo-AO system initiates a shutdown procedure.  The telescope dome closes, and the telescope moves to a parking position.  The instrument subsystems shut down, with any reference frames required taken as necessary.  Once all systems are shut down, a data archiving routine starts and compress the nightÕs data into an archive that can be transmitted off site or left in place.  All data, logs, telemetry, configuration files, and any other useful information are included in this archive.  At the completion of this process, a summary of the night of observations is compiled automatically and emailed to the appropriate people (including but not limited to the Robo-AO team); this is also be placed onto a website for external examination.

The safety of the entire system is paramount through this entire process.  At any time a violation of the safety limits is detected, the safety system takes a preset course of action to render the instrument and telescope safe from further harm.  This course includes closing the telescope dome, placing the telescope in a safe position, terminating LGS operations and shutting the laser down, placing all Robo-AO instruments in a safe state, and any other steps deemed necessary.  The key to making this work properly is identifying actions that are unsafe for the robotic system to undertake, or conditions that are unsafe.

   \begin{figure}
   \begin{center}
   \begin{tabular}{c}
   \includegraphics[height=12cm]{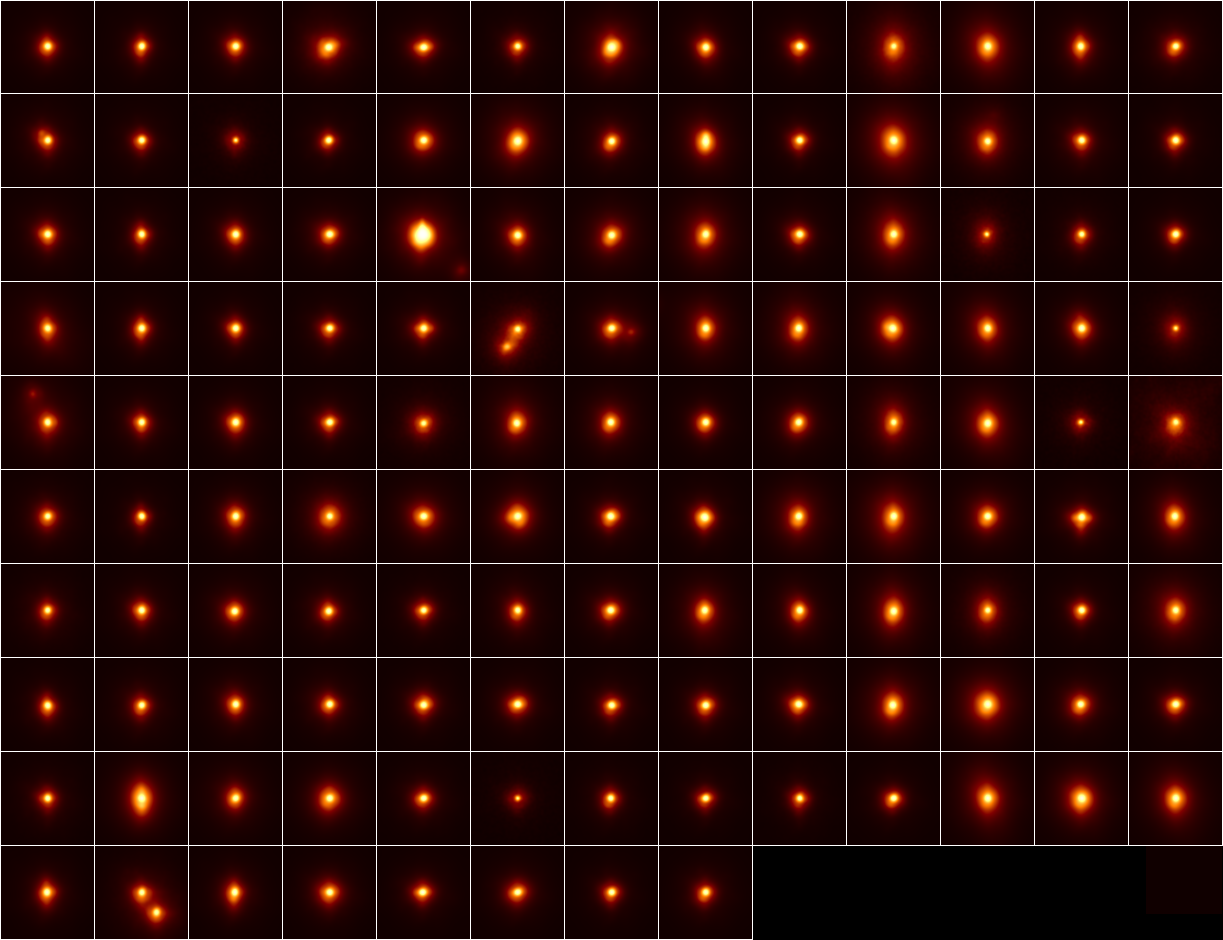}
   \end{tabular}
   \end{center}
   \caption[flowchart] 
   { \label{fig:data} Results of the first full night of fully robotic Robo-AO operations.  During this night, the instrument completed these observations without human intervention.  The data was gathered in i band, and each frame is a 90 second exposure.  Each image is a 2 arcsec square around the target; for the night of observations, seeing was 1.5-2 arcseconds.  This is a ``quick look" data analysis; further analysis will be done for the science programs.  }
   \end{figure} 

\section{PRELIMINARY PERFORMANCE RESULTS}

The Robo-AO software was developed during a three year period of extensive testing of each of the subsystems to increase their robustness and efficiency.  The entire robotic system was developed to make it run as smoothly as possible, with a targeted setup time for the AO operations of 60 seconds.  In this time, the AO system calibrates the WFS system, automatically acquires and centers the laser, configures the science instrument (i.e. set up the camera and filter wheels), and additional minor tasks required to take an observation.

The first night of full robotic operations was achieved on the night of June 17-18, 2012, when the automation system ran for a total of 7.5 hours through the 8.5 possible hours that night (30 minutes were lost to technical issues, and another 30-40 minutes were lost to blanket LGS closures by USSTRATCOM).  During this time, the system completed a total of 133 observations of stellar targets, all with an exposure time of 90 seconds; 125 of these observations are plotted in Figure~\ref{fig:data}..  The average overhead for each observation was 115 seconds; this time includes the time to slew the telescope across the sky, as well as instrument setup and operation.

An analysis of the logs shows that the instrument takes 50-60 seconds to set up the instrument for each observation, depending on the time to acquire the LGS; this time does not include the telescope slew time.  When doing multiple observations on the same target, the overhead is a few seconds required to change the filters.  A few extra seconds are used for various bookkeeping steps.  These times do not include the time required to deal with system errors, which can add considerably to the overhead time if the error is serious.

The network power switch currently in use adds 10-15 seconds to this time by itself; replacement of this piece of hardware with a fast switching system will decrease the overhead to 35-50 seconds per observation for the instrument setup.  There are many areas where the overhead time can be decreased, likely adding up to a few more seconds that will be removed from the observational overhead.  

It should be noted that setting up an adaptive optics system for observation in less than a minute is quite rapid \cite{2006PASP..118..297W}, and the next generation large telescopes all will require an automation of tasks of the same order of magnitude as the Robo-AO robotic system in order to achieve their operational requirements\cite{2010SPIE.7736E...2E}.

\section{CONCLUSION AND FUTURE WORK}

Three years of development has resulted in the production of software that operates the Robo-AO laser guide star adaptive optics and science system completely automatically; this is the first automated LGS AO system in astronomy.  During development, we were able to produce scientific results as part of the engineering tests\cite{2011arXiv1112.1701L}.  Now that the automation system is functional, the Robo-AO instrument is currently producing data as part of the initial month long scientific demonstration period on the Palomar 60-inch telescope.  Starting in August, 2012, Robo-AO will be observing for several programs allocated by the Palomar time allocation committee.  Throughout this time, the software will continue to be improved by increasing the efficiency of the system, improving the error handling, debugging, and adding new features.  The main features still to be added are the watchdog daemon system, and a queue scheduling system.  A prototype queue scheduling system has already been developed, but it requires enhancements (especially in the interaction with LGS closure windows) and integration into the observing system.

The Robo-AO was developed in a modular way to make it easy to replicate the system for other telescopes.  The Inter-University Centre for Astronomy and Astrophysics in India has partnered with Caltech to develop Robo-AO, and is in the early stages of developing a copy for the Girawali Observatory 2m telescope.  Pomona College has developed a natural guide star AO system based on Robo-AO that uses the AO control code\cite{2012AAS...22013504M}.  The hope is that this process will continue, more 1-3m telescopes will be able to operate at their diffraction limit and new scientific areas will continue to bear science fruit.

\acknowledgments     
 
The Robo-AO project is a collaboration between Caltech Optical Observatories and the Inter-University Centre for Astronomy and Astrophysics. It is partially funded by the National Science Foundation under grants AST-0906060 and AST-0960343, the Office of Naval Research under grant N00014-11-1-0903, a grant from the Mt. Cuba Foundation, and by a gift from Samuel Oschin.  We are grateful for the continued support of the Palomar Observatory staff for their ongoing support of Robo-AO on the 60-inch telescope, particularly S. Kunsman, M. Doyle, J. Henning, R. Walters, G. Van Idsinga, B. Baker, K. Dunscombe and D. Roderick.


\bibliography{report}   
\bibliographystyle{spiebib}   

\end{document}